\documentclass[%
 aip,
 sd,%
 amsmath,amssymb,
 preprint,%
author-year,%
]{revtex4-1}
\usepackage{color}
\usepackage{graphicx}
\usepackage{subfigure}
\usepackage{dcolumn}
\usepackage{bm}
\usepackage[colorlinks,
                   linkcolor=blue,
                  anchorcolor=red,
                  citecolor=blue
                  ]{hyperref}
\def\pmb#1{\setbox0=\hbox{#1}%
\kern-.025em\copy0\kern-\wd0 \kern.05em\copy0\kern-\wd0
\kern-.025em\raise.0433em\box0 }

\newcommand{\bsub}{\begin{subequations}}
\newcommand{\esub}{\end{subequations}}

\newcommand{\om}{\omega}
\newcommand{\ga}{\gamma}

\newcommand{\ep}{\epsilon}

\newcommand{\po}{\mbox{\boldmath $\omega$}}

\newcommand{\pt}{\mbox{\boldmath $\tau$}}

\newcommand{\ptau}{\mbox{\boldmath $\tau$}}

\newcommand{\ps}{\mbox{\boldmath $\sigma$}}

\newcommand{\bI}{\mathbf  I}

\newcommand{\cD}{\mathcal D}

\newcommand{\cV}{\mathcal V}

\newcommand{\pD}{\textbf{\emph{D}}}

\newcommand{\pF}{\textbf{\emph{F}}}

\newcommand{\pI}{\textbf{\emph{I}}}
\newcommand{\pJ}{\textbf{\emph{J}}}

\newcommand{\pS}{\textbf{\emph{S}}}

\newcommand{\pU}{\textbf{\emph{U}}}

\newcommand{\pP}{\textbf{\emph{P}}}

\newcommand{\pa}{\textbf{\emph{a}}}

\newcommand{\pe}{\textbf{\emph{e}}}
\newcommand{\pf}{\textbf{\emph{f}}}
\newcommand{\pgg}{\textbf{\emph{g}}}

\newcommand{\pn}{\textbf{\emph{n}}}

\newcommand{\pq}{\textbf{\emph{q}}}

\newcommand{\pu}{\textbf{\emph{u}}}

\newcommand{\pw}{\textbf{\emph{w}}}
\newcommand{\px}{\textbf{\emph{x}}}

\newcommand{\ppm}{\textbf{\emph{m}}}

\newcommand{\ptt}{\textbf{\emph{t}}}

\newcommand{\pat}{\partial}
\newcommand{\na}{\nabla}

\newcommand{\beq}{\begin{equation}}
\newcommand{\eeq}{\end{equation}}
\newcommand{\bsubeq}{\begin{subequations}}
\newcommand{\esubeq}{\end{subequations}}
\newcommand{\beqn}{\begin{eqnarray}}
\newcommand{\eeqn}{\end{eqnarray}}
\newcommand{\fr}{\frac}
\newcommand{\lb}{\label}
\newcommand{\er}{\eqref}

\begin{document}

\preprint{AIP/123-QED}

\title[Under consideration for publication in Phys. Fluids.]{Minimum-domain impulse theory for unsteady aerodynamic force}

\author{L. L. Kang$^1$, L. Q. Liu$^2$, W. D. Su$^1$ and J. Z. Wu}
\email{jzwu@coe.pku.edu.cn}
\affiliation{
$^1$State Key Laboratory for Turbulence and Complex Systems, College of Engineering, Peking University, Beijing 100871, P. R. China \\
$^2$Center for Applied Physics and Technology, College of Engineering, Peking University, Beijing 100871, China
}

\date{\today}

\begin{abstract}
We extend the impulse theory for unsteady aerodynamics, from its classic global form to finite-domain formulation then to minimum-domain form, and from incompressible to compressible flows.
For incompressible flow, the minimum-domain impulse theory raises the finding of Li and Lu (\textit{J. Fluid Mech.}, {\bf 712}: 598-613, 2012) to a theorem:
The entire force with discrete wake is completely determined by only the time rate of impulse of those vortical structures still connecting to the body, along with the Lamb-vector integral thereof that captures the contribution of all the rest disconnected vortical structures.
For compressible flow, we find that the global form in terms of the curl of momentum $\na\times (\rho \pu)$, obtained by Huang (\textit{Unsteady Vortical Aerodynamics}. Shanghai Jiaotong Univ. Press, 1994), can be generalized to having arbitrary finite domain, but the formula is cumbersome and in general $\na\times (\rho \pu)$ no longer has discrete structure and hence no minimum-domain theory exists. Nevertheless, as the measure of transverse process only, the unsteady field of vorticity $\po$ or $\rho \po$ may still have discrete wake. This leads to a {\it minimum-domain compressible vorticity-moment theory} in terms of $\rho \po$ (but it is beyond the classic concept of impulse).
These new findings and applications have been confirmed by our numerical experiments.
The results not only open an avenue to combine the theory with computation-experiment in wide applications, but also reveals a physical truth that it is no longer necessary to account for all wake vortical structures in computing the force and moment.
\end{abstract}

\pacs{Valid PACS appear here}
\keywords{Suggested keywords}
\maketitle

\section{Introduction}\label{sec.Introduction}

\begin{figure}
  \centerline{\includegraphics[width=0.4\textwidth]{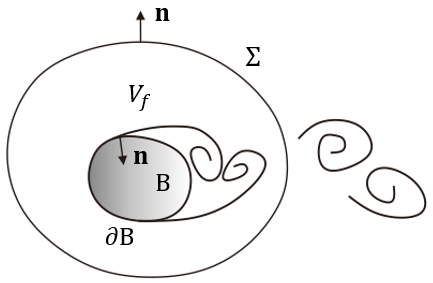}}
  \caption{Definition of notions for a moving/deforming body through the fluid in an arbitrary control volume $V =V_f+B$ bounded by $\Sigma$, which can extend to the whole free space $V_\infty$ with $\Sigma =\pat V_\infty$. The boundary of fluid $V_f$ alone is $\pat V_f =\Sigma +\pat B$.}
\label{fig.1}
\end{figure}

The theoretical prediction of total force and moment exerted on material bodies of arbitrary motion and deformation in a unbounded viscous fluid has been a very basic task of aerodynamics. Good theories should be able to shed light onto the key physical mechanisms responsible for the force and moment. Among various existing theories, special attention should be paid to the impulse theory
pioneered by \cite{Burgers1920} and later developed independently by \cite{Wu1981, Wu2005} and \cite{Lighthill1986}, among others. Take the total force as example. For an arbitrary moving/deforming body in incompressible flow, using the notation shown in Fig.~\ref{fig.1}, the theory gives\footnote{Equation \er{IF1} was given by \cite{Wu2005} and \cite{Wu2015}, which avoids calculating the virtual fluid motion inside the body as in the original form in \cite{Burgers1920} and \cite{Wu1981}, which is almost impossible when the body is deforming.}
\beq\lb{IF1}
\pF = -\fr{d\pI_{f \infty}}{dt}+\fr{1}{k}\fr{d}{dt}\int_{\pat B}\px\times (\pn\times \rho \pu)dS
\eeq
where
\beq\lb{Ifinfty}
\pI_{f\infty} \equiv  \fr{1}{k}\int_{V_{f\infty}}\px\times \rho \po dV,\ \ \ \po =\na\times \pu,
\eeq
is the vortical impulse of $V_{f\infty} =V_\infty -B$ with $k=n-1$, $n=2,3$ being the spatial dimension, where $V_\infty$ is the whole free space occupied jointly by the fluid and the body, and the second integral over material body surface $\pat B$ reflects the effect of boundary condition, which for active motion and deformation is prescribed. Here, for consistency with compressible-impulse theory to be studied later, the density $\rho$ has been included in the definition of impulse.
This formula is proven by \cite{Protas2011} to be not applicable to steady flows as a result of the slow decay of the steady-state velocity and vorticity fields as compared to the time-dependence case.

We remark that, in his publications, \cite{Wu1981, Wu2005} did not use the word and concept of impulse but called this theory {\it vorticity-moment theory}. Following \cite{Burgers1920} and most literature, we shall use the word ``impulse'' to its maximal extent. However, it turns out that although vorticity-moment theory and impulse theory refer to exactly the same formulation for incompressible flow, they differ from each other as we proceed to compressible flow (Section~\ref{sec.3}).

In addition to its incredible neatness, \er{IF1} reveals several important features of the impulse theory. Firstly, it is a direct consequence of Newton's third law, without involving the Navier-Stokes equations. Thus it has great generality, valid at any Reynolds number. Secondly, like all successful classic aerodynamic theories ever since Prandtl's (1918) vortex-force theory for steady flow, with the Kutta-Joukowski lift formula as its two-dimensional (2D) special case, the impulse theory does not require the knowledge of the pressure field. Thirdly, because its linear dependence on vorticity, the impulse can be treated as a superposition of impulses of every individual vortical structures.

These advantages strongly suggest that, once combined with modern computational fluid dynamics (CFD) and/or experimental fluid dynamics (EFD), the impulse theory may serve as a primary theoretical tool in unsteady aerodynamics, in particular in the field of biological locomotion. For example, among others, by using the impulse theory, \citet{Hamdani2000} found that during the impulsive starts of a two-dimensional wing, the large vortex at trailing-edge during fast pitching-up rotation causes a large aerodynamic force; \citet{Birch2003} examined the influence of wing-wake interactions on the production of aerodynamic forces in flapping flight; \citet{Wang2010} identified the roles of vortex rings in lift production or reduction; \cite{Kim2013} investigated vortex formation and force generation of clapping plates with various aspect ratios and stroke angles; and \cite{Andersen2016} studied the close relation between the wake patterns and transition from drag to thrust on a flapping foil.

On the other hand, the impulse theory has been generalized to viscous compressible flow by \cite{Huang1994}:
\beq\lb{impulse-Huang}
 \pF = -\fr{ d \pI^*_{f \infty } }{dt} + \fr{1}{k} \fr{d}{dt}\int_{\pat B} \px\times (\pn \times \rho \pu)dS,
\eeq
with
\beq\lb{I*0}
 \pI^*_{f \infty} \equiv \fr{1}{k} \int_{ V_{f\infty} } \px \times (\na\times \rho \pu) dV.
\eeq
But so far no application of \er{impulse-Huang} has been reported. We expect that it could be a valuable tool in the study of super-maneuvering vehicle flight.

Despite its generality and neatness as well as its relative easy to compute, however, the classic form of impulse theory has an inherent limitation. It requires calculating the entire vorticity field in $V_\infty$ or $V_{f\infty}$, or we may say that the theory is of {\it global form.} In contrast, the domain in CFD/EFD, bounded by $\Sigma$,  is always finite with some vorticity inevitably going out of $\Sigma$ shortly  after the body motion starts. This situation has made it difficult to extensively use \er{IF1} or \er{impulse-Huang} in practice. Since the first efforts to combine the impulse theory with experimental and numerical data by \cite{Birch2003} and \cite{Sun2004}, respectively, so far it has been mostly confined to dealing with a sudden-start motion or flapping wings before the body-generated vorticity escapes out of a finite domain, beyond which \er{IF1} is invalid. Therefore, it is highly desired to extend the global form of the impulse theory, both incompressible and compressible, to a finite-domain form, such that for any body motion and deformation one can always utilize the vorticity-distribution data provided by CFD/EFD to diagnose the force constituents.

Mathematically, a general impulse formulation of aerodynamic force for arbitrary finite domain has been given by \citet{Noca1997,Noca1999}.
It was also presented in \cite{Wu2015} who also proved its full equivalence to the unsteady vortex-force theory.
This general formulation also gets its wide application thanks to the boundedness and arbitrariness of integral domain.
Although the force prediction by the general formulation is accurate, however, it contains cumbersome boundary-integral terms with complicated physical meaning, making it difficult to pinpoint the dominant dynamic mechanisms responsible for the force. So the issue is whether
the formulation can be significantly simplified to a powerful theoretical-physical tool for practical applications.

Here, a key physical observation is:
unlike steady flows where the wake is always continuous, unsteady wakes behind flapping wings are often (though not always) discrete, for
which the finite-domain impulse theory can be greatly simplified, which can then clearly reveals some simple physics of crucial
importance.
A pioneering work toward this direction was made by \citet{Li2012} in a theoretical-numerical study of viscous and unsteady wake generated by flapping plates in relatively slow forward motion. There, the wake was found to be two rows of almost discrete vortex rings. The authors presented a finite-domain impulse formulation, and then found numerically that, the force of flapping plate is dominated by just the two vortices that still connect to the body.

Inspired by the work of \cite{Li2012}, in this paper we focus on proving theoretically that their numerical finding is of general significance: As long as wake vortices are discrete, the analysis domain in CFD/EFD can be minimized to a zone enclosing the body and those body-connected vortical structures. The force is solely determined by the time rate of the impulse of body-connected structures and a Lamb-vector integral thereof. The latter captures the contribution of all disconnected vortices in the wake.
This implies a {\it minimum-domain} impulse theory for discrete wake. Then, for compressible flow, we find a {\it minumum-domain compressible vorticity-moment theory}.

The organization of the paper is as follows. We study the incompressible impulse theory in an arbitrary finite domain first in Section~\ref{sec.2}, including a sharpened proof of the classic global theory \er{IF1}. We then prove that once the wake has discrete compact vortical structures, the finite-domain impulse theory can be greatly simplified if the outer boundary of analysis domain does not cut any compact vortices (no-cut condition), and the smallest one of such domain is optimal. In Section~\ref{sec.3} we proceed to a unified general impulse formulation for the force for compressible unsteady flow in an arbitrary finite domain, which is exact but cumbersome, and has no way to be simplified because $\na\times (\rho\pu)$ in \er{I*0} does not have discrete wake. We thus propose to an alternative force formula in terms of $\rho \po$, that goes beyond the framework of impulse but still stays within the realm of vorticity-moment theory, where the no-cut condition is as simple as that for incompressible flow. These theoretical findings are tested by some numerical examples presented in Section~\ref{sec.4}. Concluding remarks are given in Section~\ref{sec.5}. For neatness, we shall consider the force theory only, because the moment formulas can be easily derived in a closely similar way and do not need to be discussed separately.

\section{Incompressible force theory by finite-domain impulse}\label{sec.2}

For clarity, this section is confined to incompressible flow with $\rho = \rm const$. After reviewing the concept of impulse and relating it to the general standard total-force formula to set a common basis for later development, we first re-derive the classic neat formula \er{IF1} but in a finite-domain formulation, and then go on to examine the formulation in arbitrary finite-domain.
The fact that many unsteady problems like jellyfish propulsion and birds flapping wings involve discrete wakes, motivates us to the desired minimum-domain theory.

\subsection{From impulse to aerodynamic force}\lb{sec.2.1}

The original concept of impulse introduced by \cite{Thomson1869},  see also \cite{Lamb1932} and \cite{Batchelor1967}, was to bypass the non-compact part of the total momentum that has poor divergence behavior as $r \to \infty$. Actually, the concept and definition of vortical impulse can well be introduced in different ways. For example, by using a gauge transformation, \cite{Kuzmin1983} introduced directly an impulse-density field $\pq =\pu+\na \varphi$ for incompressible flow and defined the gauge field $\varphi$ in such a way to eliminate the pressure.

Similar to but more straightforward than Kuzmin's (1983) approach, we can use a derivative-moment transformation (DMT), see \er{DMTa} in Appendix \ref{app.A}, to split the total momentum in a fluid domain $V_f$ to
\beq\lb{C-NC}
\pP_f \equiv \int_{V_f}\rho \pu dV = \pI_f-\pS_f,
\eeq
where
\bsubeq\lb{IS}
\beqn
\pI_f &\equiv & \fr{1}{k}\int_{V_f}\px\times \rho\po dV,\lb{I}\\
\pS_f &\equiv& \fr{1}{k}\int_{\pat V_f}\px\times (\pn\times \rho \pu)dS = \pS_\Sigma +\pS_B,\lb{S}
\eeqn
\esubeq
where the subscripts $\Sigma$ and $B$ denote surface integrals over external and internal boundaries of $V_f$, respectively.

The central issue in the impulse theory of aerodynamic force is to express the standard total-force formulas of field form by the time rate of impulse.  Let $\pat B$ and $\Sigma$ be material surfaces. Then the total force acting on the body is
\beq\lb{Ffielda}
 \pF = -\fr{d}{dt}\int_{\cV_f} \rho \pu dV + \int_{\Sigma} (-p \pn +\ptau) dS,
 \eeq
where $\ptau =\mu\po\times \pn$ is the shear stress. Therefore, for incompressible flow, substituting \er{C-NC} into \er{Ffielda} yields
\beq\lb{impulseforce}
\pF= -\fr{d\pI_f}{dt} +\fr{d\pS_B}{dt}+\fr{d\pS_{\Sigma}}{dt}+\int_{\Sigma}(-p \pn + \ptau)dS,
\eeq
which serves as a common basis of following developments. For latter use, kinematic content of $d\pS_f/dt$ and the dynamic content of $d\pI_f/dt$ are given in Appendices \ref{app.B} and \ref{app.C}, respectively, including both incompressible and compressible flows.

\subsection{Finite domain containing entire vorticity field}\lb{sec.2.2}

Owing to the physical compactness of vorticity field, $\pI_f$ must remain finite. In contrast, $\pS_{\Sigma}$ represents a non-compact part of $\pP_f$ and is only conditionally convergent as $|\px|\to \infty$. The essential requirement of the classic global impulse theory is that the domain $V =V_f+B$, which may well be finite, should contain the entire compact vorticity field. Thus we impose a condition
\beq\lb{irr-cond}
 \pu = \na \phi \quad {\rm at}\ \Sigma\ {\rm and\ outside}\ V.
\eeq
In this case,  the splitting \er{C-NC} is very close to an inherent longitudinal-transverse (L-T) decomposition of the total momentum, where the longitudinal part (characterised by curl-free field) is $\pS_{\Sigma}$ and transverse part (characterised by solenoidal field) is $\pI_f$.

 Recall that $D(d\pS)/Dt = -d\pS\cdot (\na \pu)^T$ \citep[][p~132]{Batchelor1967}, where $d\pS=\pn dS$ is the material surface element, there is
\beq\lb{dS-dt}
 \fr{d\pS_{\Sigma}}{dt} = -\rho \int_\Sigma \left( \fr{D\phi}{Dt}\pn - \phi\pn\cdot \na \na \phi \right) dS.
\eeq
Since the flow outside $\cV$ is irrotational, the $\Sigma$-integral of $\phi\pn\cdot \na\na\phi$ in \er{dS-dt}  is equivalent to a volume integral over a potential-flow domain $\cV_e$ bounded internally by $\Sigma$ (with $\hat{\pn}=-\pn$ being the unit normal out of $\cV_e$) and externally by a larger control surface $\Sigma'$ that may approach $V_{\infty}$. Then there is
\beq\lb{IntVe}
 \int_{\pat \cV_e} \phi \hat{\pn} \cdot \na \na \phi dS = \int_{\cV_e} \left( \fr{1}{2} \na |\na\phi|^2
 + \phi \na \na^2 \phi \right) dV = \int_{\pat \cV_e} \fr{1}{2}|\na\phi|^2 \hat{\pn} dS
\eeq
due to $\na^2\phi =0$ in $\cV_e$. Here, $\pat \cV_e =\Sigma + \Sigma'$, but owing to algebraic decay of $\phi$, the boundary integrals over $\Sigma'$ in \er{IntVe} are negligible. Thus, \er{dS-dt} is reduced to
\beq\lb{Ber}
 \fr{d\pS_{\Sigma}}{dt} = - \rho \int_\Sigma \left( \fr{D\phi}{Dt} - \fr{1}{2} |\na\phi|^2 \right) \pn dS = \int_\Sigma p\pn dS
\eeq
by the Bernoulli equation in Lagrangian description. Therefore, the third and fourth terms in \er{impulseforce} are cancelled and we obtain \er{IF1}, which is equivalent to the original classic impulse theory \citep{Wu1981}, but has the advantage of finite domain over the latter.

\subsection{Impulse formulation in arbitrary finite domain}\label{sec.2.3}

We now drop the condition \er{irr-cond} and return to \er{impulseforce}. To reveal various physical roots of the difference between the total momentum and impulse in a generic finite domain, we shift $d/dt$ of the third term in \er{impulseforce} into the integral, of which the algebra is lengthy and given in Appendix \ref{app.C}. For a material $\Sigma$, substituting \er{B8} into \er{impulseforce}, we obtain the impulse formulation in arbitrary material finite domain
\beq\lb{F2}
\pF= -\fr{d\pI_f}{dt} -\int_{\cV_f}\rho\po\times\pu dV+\pF_{\pat B}
+\fr{1}{k}\int_\Sigma\px\times \rho\pu \omega_n dS+ \pF_\Sigma ,
\eeq
where
\beq\lb{FB}
\pF_{\pat B} \equiv \fr{1}{k}\int_{\pat B}\px\times (\pn\times \rho\pa)dS +\fr{1}{k}\int_{\pat B}\px\times\rho\pu \omega_n dS
\eeq
and
\beq\lb{FS}
\pF_{\Sigma} \equiv \fr{1}{k}\int_{\Sigma}(\px\times \rho\ps +\ptau)dS
\eeq
is a viscous effect at $\Sigma$ with $\ps=\nu\pat \po/\pat n$ being the vorticity diffusive flux.

Some terms of \er{F2} need to be explained. First, the vortex-force term, i.e., the Lamb-vector integral, is the major inevitable addition to \er{IF1}. As \cite{Saffman1992} explained, this term is the effect of the vortical flow outside $\cV_f$ (including the virtual fluid in $B$) on the force. Physically, one may replace $\po\times \pu$ in this term by $\po\times \na\phi_e$, where $\na\phi_e$ is the potential velocity induced by vorticity outside $\cV_f$ --- although this is inconvenient for calculation --- and hence when the outer vorticity is far away from $\cV_f$ then this vortex-force term may be neglected.

On the other hand, $\pF_{\pat B}$ represents the explicit effect of body motion and deformation, which for active motion/deformation is a prescribed integral due to the adherence of $\pu, \pa$, and $\om_n$ at $\pat B$, and is completely independent of the flow field. Thus, $\pF_{\pat B}$ serves as a {\it driving mechanism} of the flow field, which is in contrast to the rest terms of \er{F2} that represent the {\it fluid reaction} to the body's driving. Alternative to \er{FB}, we may also write
\beq\lb{FB+}
\pF_{\pat B} = \fr{d\pS_B}{dt} -\int_B\rho\po\times\pu dV=\fr{d\pS_B}{dt}+\int_{\cV_{f\infty}}\rho\po\times\pu dV.
\eeq
Compared to the body-surface integral in \er{IF1}, we see that once some vorticity is out of $\cV_{f\infty}$ the body-surface influence can no longer be described by $d\pS_B/dt$ alone, unless the body volume is negligible as assumed by \cite{Li2012}. Since $\pn\times \pa \equiv \ps_a$ in \er{FB} is the vorticity creation rate (or boundary vorticity flux) caused by tangent acceleration of the wall, of which an examination tells how the opposite vorticity fields at upper and lower surfaces of the wing, say, is generated --- a crucial information in understanding the kinetics of the entire vortical wake pattern but not seen in \er{FB+}, below we shall use \er{FB} exclusively.

The $\Sigma$-integrals contain two terms: one is $\pF_\Sigma$ which represents the viscous effect and can be ignored at large Reynolds number, and the other can not be ignored if the outer boundary cuts considerable vortices.

Equation \er{F2} is exact and general for incompressible flow. Compared with the general impulse theory proposed by \cite{Noca1997, Noca1999}, \er{F2} (along with Appendices \ref{app.B}) has neater form and clearer physical meaning, of which the compressible form is shown in Subsection~\ref{sec.3.1}.
 It is fully equivalent to the unsteady vortex-force theory \citep{Wu2015} as can be shown by using the Reynolds transport theorem, so it can also be used in steady flow, but here we are more concerned with the unsteady flow problem.
 Due to the appearance of multiple $\Sigma$-integrals, \er{F2} can hardly catch the key physical mechanism for producing force. Unlike $\pF_{\pat B}$ which is inevitable for moving-deforming body, these $\Sigma$-integrals are no more than a necessary artifice to express the force of the body in externally unbounded fluid by the flow data in finite domain. Their appearance makes the characteristic neatness of impulse theory completely lost. A natural step to simplify \er{F2} is evidently to remove the $\Sigma$-integrals. We do this in the next subsection.
\subsection{Minimum-domain impulse theory for flow with discrete wake structures}\label{sec.2.4}

Unlike steady flow (viewed in the inertial frame of reference fixed to the body) where the wake vorticity field must be continuous, a remarkable feature of unsteady flow is that in many cases, such as birds flapping wings, fish propulsion and flow around blunt body, the wake consists of discrete or compact vortical structures. This fact has made it possible for approximating the wake vortex street by point vortices as \cite{Karman1911} did and later by arrays of vortex patches as reviewed by \cite{Saffman1992}. Although in real unsteady flow the vortices shed into the wake may still be connected by vortex sheets, compared to concentrated vortices formed by vortex-sheet rolling up, the vorticity in the sheets is much weaker, and the intersections of $\Sigma$ and the sheets are  too small to have appreciable effect on the force. Therefore, we can choose special $\Sigma$ to avoid cutting any discrete vortical structures. This amounts to a condition much weaker than
\er{irr-cond}:
\beq\lb{no-cut}
\po={\bf 0}\ \ {\rm at\ and\ near}\ {\Sigma}.
 \eeq
Then we call an outer domain boundary satisfying \er{no-cut} a {\it good $\Sigma$} and the fluid domain bounded by a good $\Sigma$ a {\it good $\cV_f$}. By \er{no-cut} all surface integrals over $\Sigma$ in \er{F2} disappear.

There can be more than one good $\Sigma$'s when the wake has many discrete vortical structures. But for any compact vortical domain, say $\cV_{f\rm wk}$, bounded by a good $\Sigma$, by \er{A4} and \er{no-cut} the fluid exerts no force to the body apperantly:
\beq\lb{fi}
 \pF_{f{\rm wk}} =-\fr{d\pI_{f{\rm wk}}}{dt}-\int_{\cV_{f{\rm wk}}}\po\times \pu dV ={\bf 0},
\eeq
no matter how complex the vortical structures could be in $\cV_{f{\rm wk}}$. For example, they could even be some bundles of mutually tangled vortices. The same conclusion was made by \cite{Saffman1992} for inviscid flow with $\pat\cV_f$ satisfying $\po\cdot\pn=\bf 0$. In fact, \er{fi} implies that $d\pI_{f\rm wk}/dt$ equals to the Lamb-vector integral over $V_\infty-\cV_{f\rm{wk}}$.

This being the case, from a good $\cV_f$ we can always identify a ``body-connected zone'' $\cV_{f{\rm con}}$ bounded by a good $\Sigma$, just like that sketched in Fig.~\ref{fig.1}, which contains all vortical structures still connecting to the body, including attached boundary layers, separated shear layers and rolled-up vortices thereby; then the rest of $\cV_f$ will certainly belong to $\cV_{f{\rm wk}}$ and does not contribute to \er{F2}. Therefore, we obtain the desired minimum-domain force formula by impulse:
\beq\lb{F6}
 \pF = -\fr{d\pI_{f{\rm con}}}{dt}-\int_{\cV_{f{\rm con}}} \rho\po \times \pu dV  + \pF_{\pat B},
\eeq
where $\pF_{\pat B}$ represents the driving mechanism and the first two terms represent the fluid reaction to the body. Actually, upon dividing the whole vortical domain $\cV_{f\infty}$ into $\cV_{f\rm con}$ and $\cV_{f\rm wk}$, \er{F6} can be more directly derived by substituting \er{fi} and \er{FB+} into \er{IF1}. This result reveals a remarkable physical truth, worth being stated as  a theorem:

{\bf Theorem.} {\sl When a body moves and deforms arbitrarily in an externally unbounded viscous fluid, if the wake consists of a set of discrete compact structures, then except the driving mechanism $\pF_{\pat B}$, the fluid reaction to the body's motion exerts a force that can be solely determined by the rate of change of the impulse of the vortical structure still connecting to the body, along with the associated Lamb-vector integral thereof.}

This theorem proves that the numerical finding of \cite{Li2012} represents a common phenomenon for any unsteady flow with discrete wake structure.\footnote{In the case studied by \cite{Li2012}, after being disconnected from the flapping wing, the wake vortex rings are sufficiently far from each other so that their mutual induction is negligible. Consequently, both terms of \er{fi} vanish simultaneously.}

It should be stressed immediately that the above theorem by no means implies that the body-connected vortical structure can be an isolated existence. It always coexists and interact all wake vortices, either kinetically or kinematically. The point is: Once disconnected from the body, any wake vortices, no matter how many, are irrelevant to the total force (and moment). Counting on all these vortices as one usually does is perfectly OK, as will be numerically confirmed in Section~\ref{sec.4}; but that counting is redundant. This situation is just like the fact that one only needs the pressure $p$ and friction $\ptau$ over the body-surface for calculating the total force, although the $(p,\ptau)$-field on $\pat B$ is not an isolated existence but has to be determined by the entire flow in space.

\section{Compressible force theory by finite-domain impulse}\label{sec.3}

Apparently, the generalization of the above incompressible impulse theory to compressible flow is straightforward. Here we use short-hand notations for the momentum of unit volume and its divergence and curl:
\beq\lb{m-om*}
 \ppm \equiv \rho \pu,\ \ \ \vartheta^* \equiv \na\cdot \ppm, \ \ \po^* \equiv \na\times \ppm.
\eeq
Since we have argued that the impulse can be directly identified by using the DMT identity \er{DMTa} to split the total momentum, it is evident that the natural extension of impulse $\pI_f$ can only be
\beq\lb{I*}
 \pI^*_f \equiv \fr{1}{k}\int_{V_f}\px\times \po^*dV,
\eeq
as used in \er{impulse-Huang}. Note that Kuzmin's (1983) gauge-transformation approach has also led \cite{Shivamoggi2010} to identify the impulse $\pI^*_\infty$ as defined by \er{I*0} for inviscid, compressible, and unbounded flow in free space $V_\infty$, who confirmed its suitability in several aspects as the counterparts of those for incompressible flow \citep{Batchelor1967}.

Starting from $\pI^*_f$, we follow the same argument of Subsection~\ref{sec.2.1}.
By \er{DMTa} we have
 \beq\lb{P}
\pP^*_f \equiv\int_{V_f}\ppm dV = \pI^*_f-\pS^*_f,
\eeq
where $\pS^*_f =\pS^*_B+\pS^*_{\Sigma}$ is the same as $\pS_f$ defined in \er{S}, where the asterisk just reminds the variable density. For a fluid occupying the entire free space $V_{f\infty}$, $\pS^*_{\Sigma}$ vanishes due to the exponential decay of disturbances in compressible flow as proved recently by \cite{Liu2016} and \cite{Liu2017}. Then what remains in \er{P} is $\pP^*_{f\infty} =\pI^*_{f\infty}-\pS^*_B$, from which the global force formulas \er{impulse-Huang} easily follows. Below we focus on finite-domain formulation.

\subsection{Impulse formulation in arbitrary finite domain and its simplification}\label{sec.3.1}

Along with $d\pI^*_f/dt$,  in a force formula in terms of a finite domain we anticipate the appearance of a ``dynamic'' Lamb-vector (D-Lamb vector for short) integral (vortex force) as it did in \er{F6} for incompressible theory. A natural choice of D-Lamb vector is $\po^*\times \pu$. In addition to this, in developing a longitudinal-transverse (L-T) force theory,   \cite{Liu2014-jfm, Liu2014-fdr} have identified a few different Lamb-vector-like dynamic vectors. One of these is denoted by
\beq\lb{f}
 \rho \pf \equiv \na\cdot (\ppm\pu-K\bI),
\eeq
where $K=\fr{1}{2}\rho |\pu|^2$. $\rho \pf$ and $\po^*\times \pu$ are related by
\begin{equation}\lb{f-vor}
 \rho \pf = \po^* \times \pu + \vartheta \ppm + K \na \ln \rho.
\end{equation}
 Now the total force acting on the body is
\beq\lb{Ffieldb}
 \pF = -\fr{d}{dt}\int_{\cV_f} \rho \pu dV + \int_{\Sigma} (-\Pi \pn +\ptau) dS,
 \eeq
where $\Pi = p - \mu_\theta \vartheta$, $\mu_{\theta}=\lambda+2\mu$, with $\lambda$ and $\mu$ being named the second viscosity and shear viscosity. Then, by substituting \er{P} and \er{B7} into \er{Ffieldb} we obtain a general compressible impulse formulation ($\po^*$-formulation for short) for the total force in terms of flow data in arbitrary finite domain:
\bsubeq\lb{IF6}
\beqn\lb{IF6a}
 \pF &=& -\fr{d \pI^*_f}{dt} - \int_{\cV_f} \rho \pf dV-\fr{1}{k} \int_{\pat \cV_f}\px\times (\pn \times  \rho \pf ) dS\nonumber\\
  &&+\fr{1}{k}\int_{\Sigma}\px \times \po^* u_n dS +\pF_\Sigma+ \pF^*_{\pat B},
\eeqn
where $\pF_\Sigma$ is defined by \er{FS} and
\beq\lb{IF6b}
 \pF^*_{\pat B}=\int_{\pat B}\px\times(\pn\times\rho\pa)dS + \fr{1}{k}\int_{\pat B}\px \times \po^* u_n dS.
\eeq
\esubeq
By using the Reynolds transport theorem, it can also be proved that \er{IF6} is equivalent to the L-T force theory of \cite{Liu2014-jfm} in terms of unsteady and compressible vortex-force formulation. We mention that, by a reasoning similar to that in Subsection~\ref{sec.2.3},  \er{impulse-Huang} can be recovered from \er{IF6} by imposing a condition that $\po^*={\bf 0}$ outside the domain $\cV$, which is a compressible counterpart of \er{irr-cond}.

Compared to the arbitrary finite-domain formulation \er{F2} for incompressible flow, the $\Sigma$-integrals in \er{IF6} are even more cumbersome due to the complex constituents of $\rho\pf$ therein, implying that a bigger effort is needed to simplify the formulation to a practically usable one.

To this end, we first remove the $\Sigma$-integrals of quantities in $\rho \pf$ other than terms containing $\po^*$. This is not only a preparation for choosing special $\Sigma$ satisfying a no-cut condition similar to \er{no-cut} but also eases numerical calculations of these integrals. As observed by \cite{Mele2014} in their numerical test of a corresponding vortex-force formula, the $\Sigma$-integral of $\rho \pf$ in \er{IF6} contains terms $\vartheta \ppm$ and $K\na \ln \rho$ which are both discontinuous across shocks and may damage the numerical accuracy. \cite{Mele2014} pointed out that this trouble can be eliminated by using \er{DMTa} to cast the domain and boundary integrals of the unwanted terms to a single domain integral of their curl. Now, since $K\na \ln \rho =(H-h)\na \rho$, where $h$ is the enthalpy and $H =h +|\pu|^2/2$ the total enthalpy, we set
\beqn
 k\pw^* &\equiv& \na\times (\vartheta\ppm + K\na\ln \rho) \nonumber\\
 &=& \vartheta\po^*+\na\vartheta\times \ppm +\na H\times \na\rho -\fr{\ga\rho}{\ga-1}\na T\times \na s.
\eeqn
Then \er{IF6a} is cast to a form with neater $\Sigma$-integrals:
\beq\lb{F*}
 \pF = -\fr{d \pI^*_f}{dt} - \int_{\cV_f} (\po^*\times \pu+\px\times \pw^*) dV
 +\fr{1}{k}\int_\Sigma(\px\times \pu\om^*_n)dS +\pF_{\pat B}^*+ \pF_\Sigma.
 \eeq
Now the remaining $\Sigma$-integrals of $\po^*$ would be removed, if we could find good $\Sigma$'s satisfying
\beq\lb{no-cut*}
 \po^*={\bf 0}\ \ {\rm at\ and\ near}\ \ \Sigma,
\eeq
a revision of \er{no-cut}. Then \er{F*} would become
\beq\lb{F*-nocut}
 \pF = -\fr{d \pI^*_f}{dt} - \int_{\cV_f} (\po^*\times \pu+\px\times \pw^*) dV  +\pF_{\pat B}^* + \pF_\Sigma,
 \eeq
which is a compressible counterpart of \er{F2}. But \er{no-cut} does not implies $\po ={\bf 0}$ at $\Sigma$ as well, so $\pF_\Sigma$ remains. Nevertheless, at large Reynolds numbers $\pF_\Sigma$ decays rapidly away from the body \citep{Wu2007}, and at a $\Sigma$ of moderate size it is negligible.
\subsection{Beyond the impulse: A compressible vorticity-moment theory}\label{sec.3.3}
Evidently, whether the simplified force formula \er{F*-nocut} is applicable depends on the feasibility of finding special boundary $\Sigma$ that satisfies \er{no-cut*}, where
\beq\lb{oo}
 \po^* =\rho \po +\na\rho\times \pu =\rho \po -\ppm\times \na \ln \rho.
\eeq
Unfortunately, by careful analysis, we conclude that it is very difficult (if not impossible) to find a good $\Sigma$ satisfying \er{no-cut*}. The main trouble is the widespread field $\nabla\rho\times\pu$, of which some features will be demonstrated in Subsection~\ref{sec.4.3}, see Fig.~\ref{fig.4-1} below. Therefore, we leave the $\po^*$-formulation and turn to a new vector integral
\beq\lb{Irho}
 \pI_{\rho f}\equiv \fr{1}{k}\int_{\cV_f}\px\times(\rho\po)dV
\eeq
instead of $\pI^*_f$. The difference between $\pI^*_f$ and $\pI_{\rho f}$ is that the former contains certain effects of longitudinal process and hence beyond the category of vorticity-moment theory, while the latter captures the transverse process only that is always measured by vorticity, which is the very meaning of vorticity-moment theory as named by \cite{Wu1981, Wu2005}.

Then, following exactly the same algebra as used in Subsection~\ref{sec.3.1}, including removing those unwanted $\Sigma$-integrals, we find
\beqn\lb{F-rho}
 \pF &=& - \fr{d\pI_{\rho f}}{dt} - \int_{\cV_f} (\px \times \pw_\rho + \rho \po \times \pu) dV \nonumber + \pF_{\pat B}\\
 & & + \pF_\Sigma+ \fr{1}{k}\int_{\Sigma}\px\times \rho \pu \om_n dS ,
\eeqn
in which all terms have the same form as for incompressible flow but the integral of $\px\times \pw_\rho$, with $\pw_\rho$ being defined as
\beq\lb{w-rho}
 k \pw_\rho \equiv \po\vartheta^* -\na \rho \times (\po \times \pu) + \rho \na T\times \na s.
\eeq
This is a synthetic compressible effect confined in boundary layers, vortical wake, and shock waves. Then by choosing good $\Sigma$ satisfying the incompressible no-cut condition \er{no-cut}, both surface integrals on $\Sigma$ vanish at once.

Moreover, similar to \er{fi}, by \er{no-cut} and \er{A4}, any compact fluid body $\cV_{f\mathrm{wk}}$ also exerts no force to the body apparently:
\beq\lb{frhoi}
 \pF_{\rho f \rm wk}=-\fr{d \pI_{\rho f\mathrm{wk}}}{dt} - \int_{\cV_{f\mathrm{wk}}} ( \px \times \pw_\rho + \rho \po \times \pu ) dV=\textbf{0}.
\eeq
This simplifies \er{F-rho} to the desired {\it minimum-domain} compressible vorticity-moment theory:
\beq\lb{CF6}
 \pF = -\fr{d\pI_{\rho f{\rm con}}}{dt}-\int_{\cV_{\rho f{\rm con}}} ( \px \times \pw_\rho + \rho \po \times \pu ) dV \\
 +\pF_{\pat B}.
\eeq
Evidently, the theorem stated in Subsection~\ref{sec.2.4} can be extended to compressible flow, with some phrases being modified.

\section{Numerical study and physical discussions}\label{sec.4}

This section presents our numerical simulations of unsteady viscous flows around a two-dimensional elliptic airfoil to check the theoretical findings stated in Section~\ref{sec.2} and Section~\ref{sec.3}, by which we can diagnose the influence of the vortical structures to the body force. The wing starts motion impulsively at $t=0$ with angle attack $\alpha=80^{\circ}$, and then keeps moving at a constant velocity $\pU = -U\pe_x$ (from right to left).

\subsection{Numerical method and validation}\label{sec.4.1}

The OpenCFD-EC2D-1.5.4 program developed by Professor X.L. Li of Chinese Academy of Sciences is employed to solve the Navier-Stokes equations. In this open-source software, these equations are solved by a finite-volume method, with the convective terms discretized by a third-order WENO scheme and the viscous terms by a second-order central difference. For temporal terms a first-order LU-SGS (lower-upper symmetric-Gauss-Seidel) method is used with the dimensionless time step $\Delta t=t/T=0.001$. We assume constant $\mu$  and $Re=1000$, so no turbulence model is used. Quantities are made dimensionless by the chord length $L$ and speed $U$ of the airfoil, the characteristic time $T=L/U$, and by density $\rho_0$ and sound speed $c$ of uniform incoming flow. The initial condition is set as the free-stream quantities. The pressure far-field boundary condition is used on the outer boundary. No-slip and adiabatic conditions are applied on the airfoil surface.

\begin{figure}
  \centerline{\includegraphics[width=0.9\textwidth]{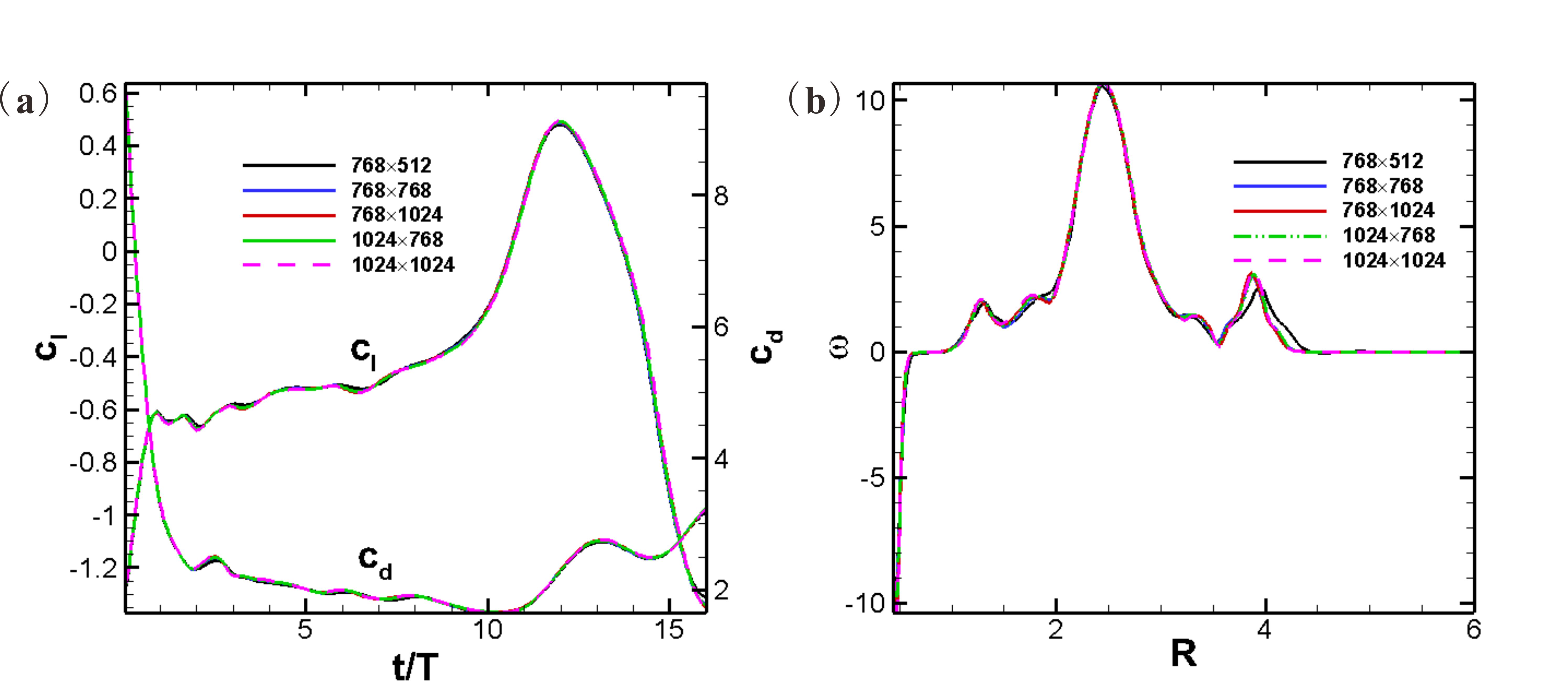}}
  \caption{(a) The lift and drag coefficients on an impulsively started elliptical airfoil with an angle of attack $80^{\circ},  M=0.4, Re=1000$; (b) The vorticity distribution in the wake along the mesh line $i=1$ at $t/T=10$.}
\label{fig.2}
\end{figure}

In this study, our computation used an O-grid with $i,j$ as the node numbers along the radial and azimuthal directions, respectively, where $i=1$ is along the semi-principle axis in the wake. Confocal ellipses and hyperbolae correspond to constant $i$ and $j$, respectively. The thickness ratio of the elliptic airfoil is $0.4$. The mesh height nearest to the airfoil surface is $0.001L$. The radius of computational domain is $R=100L$. To valid the method, convergence check with $ M=0.4, \ \alpha=80^{\circ}$ was carried out to assess the effect of mesh density. The time-dependence of lift coefficient $c_l$ and drag coefficient $c_d$ of the airfoil after the impulsive start calculated by the standard surface-stress integral,
\beq\lb{standard}
\pF=-\int_{\pat B}(-\Pi\pn+\pt)dS,
\eeq
is shown in Fig.~\ref{fig.2}(a). Figure \ref{fig.2}(b) illustrates the instantaneous vorticity distribution in the wake along the mesh line $i=1$ at $t/T=10$. The consistence of the results obtained by different grid resolutions shows that our numerical solutions are convergent.
The results given below are calculated with the number of mesh points $1024\times 1024$ at $Re =1000$.

\subsection{Incompressible flow}\label{sec.4.2}

 In Subsection~\ref{sec.2.4} we derived the arbitrary-domain incompressible impulse theory \er{F2} and the minimum-domain incompressible impulse theory \er{F6} for good $\Sigma$, where the no-cut condition \er{no-cut} is assumed. We also proved \er{fi}, that in any compact vortical domain the time rate of impulse plus the Lamb-vector integral vanishes. We now test these conclusions and analyse the influence of the vortical structures to the body force.

\begin{figure}
  \centerline{\includegraphics[width=0.9\textwidth]{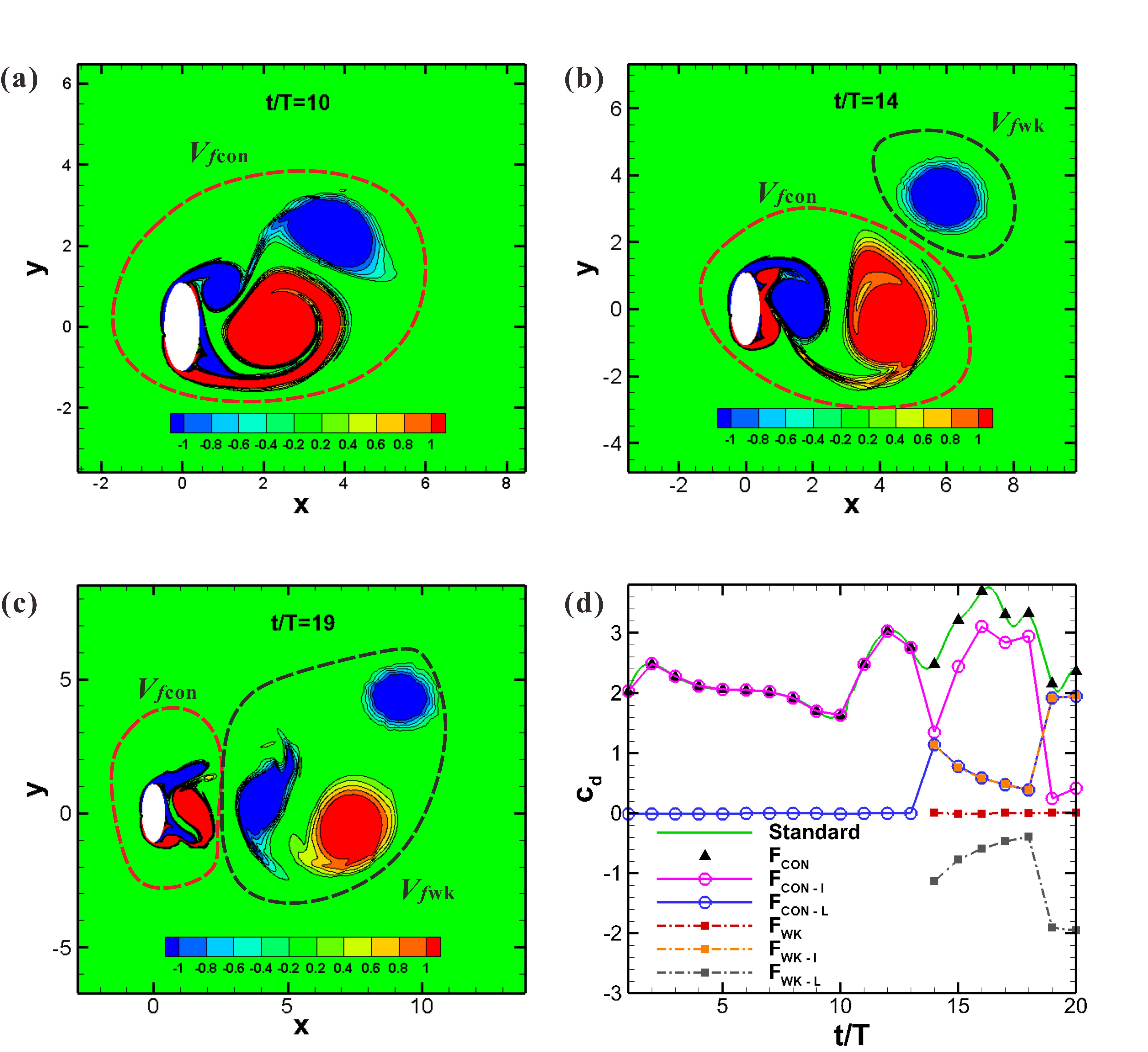}}
  \caption{(a, b, c) Instantaneous vortical structures of incompressible flow around an elliptical airfoil with $Re=1000$ at $t/T=10, 14$ and $19$, respectively. The magnitude of the vortex core in the wake is of $O(10)(3<|\po|<30)$. (d) Time dependence of the drag coefficient $c_d$, calculated by \er{fi} (${\rm F_{WK}}$), \er{F6} (${\rm F_{CON}}$) and \er{standard} (${\rm Standard}$), where the line legend marked by $I$ and $L$ represent the force contributed by -$d\pI/dt$ and the integral of $-\rho\po\times \pu$ in \er{fi} and \er{F6}, respectively.}
\label{fig.3}
\end{figure}

Figure~\ref{fig.3} (a, b, c) shows the instantaneous vorticity contours at different times. The vorticity magnitude of every vortex core in the wake is of $O(10)$, and the vorticity contour is used to highlight the compact vortical structure.
$\cV_{f\mathrm{con}}$ and $\cV_{f\mathrm{wk}}$ denote the minimum discrete vortical domain enclosing the body and compact vortical domain in the wake, respectively. The boundaries of $\cV_{f\mathrm{con}}$ and $\cV_{f\mathrm{wk}}$ are chosen to avoid cutting any compact vortices and move with the structures during the flow evolution. We mention here that there is no principle difficulty to write a subroutine to make boundary selection (preferably in terms of Lagrangian motion), but the present paper is more focused on the physics. In this and all following numerical cases, $\cV_{f\rm{con}}$ is updated to the minimum one once new disconnected vortices  are therein, although it does not matter if $\cV_{f\mathrm{con}}$ contains some extra compact vortices.
Here we set the wake domain $\cV_{f\mathrm{wk}}$ to contain all disconnected vortices.

To test \er{F6} and \er{fi}, we calculated the time-dependent drag coefficient $c_d$ of the  airfoil contributed by the vortical structures in $\cV_{f\rm con}$ and $\cV_{f\rm wk}$, respectively, as shown in Fig.~\ref{fig.3}(d), indicating that the minimum-domain impulse theory \er{F6} predicts the force very well. When $\Sigma$ cuts vortices in the wake,  it fails, but the arbitrary-domain formula \er{F2} still works as an accurate prediction (figure not shown). Some physical explanations are given below.

 When the whole wake vortical structure is connected with the body as shown in Fig.~\ref{fig.3}(a), $\cV_{f\rm{con}}$ contains all the vortices. Thus, there is $\int_{\cV_{f\rm{con}}}\po\times\pu dV=\bf 0$ before $t/T=14$ as shown in Fig.~\ref{fig.3}(d) and \er{F6} reduces to \er{IF1}.
This result confirms the aforementioned essential condition for the classic global impulse theory, i.e., $\cV_f$ contains the entire compact vorticity field rather than the entire velocity field.
When the first "fall-off'' vortex appears at about $t/T=14$, see Fig.~\ref{fig.3}(b), $\cV_{f\rm{con}}$ is narrowed and the Lamb-vector integral term starts to contribute to the force.
By \er{fi}, the compact vortex in the wake apparently has no net contribution to the total force as shown in Fig.~\ref{fig.3}(d), which confirms our theoretical conclusion. And the Lamb-vector integral in $\cV_{f\rm{con}}$ equals the time rate of the disconnected vortical structures.
$\cV_{f\rm{wk}}$ contains the first "fall-off" vortex from $t/T=14$ until $\cV_{f\rm{con}}$ is narrowed again at $t/T=19$ as shown in Fig.~\ref{fig.3}(c). From the viewpoint of the classic global form, the force contributed by the first "fall-off" vortex will decrease as moving backward, as shown by the orange line in Fig.~\ref{fig.3}(d) during $t/T=14\sim 18$, which equals to the Lamb-vector integral in $\cV_{f\rm{con}}$.
This observation is of great significant for CFD/EFD that the far-field influence can alway be captured by the near-field Lamb-vector integral.
\subsection{Compressible flow}\label{sec.4.3}

\begin{figure}
  \centerline{\includegraphics[width=0.9\textwidth]{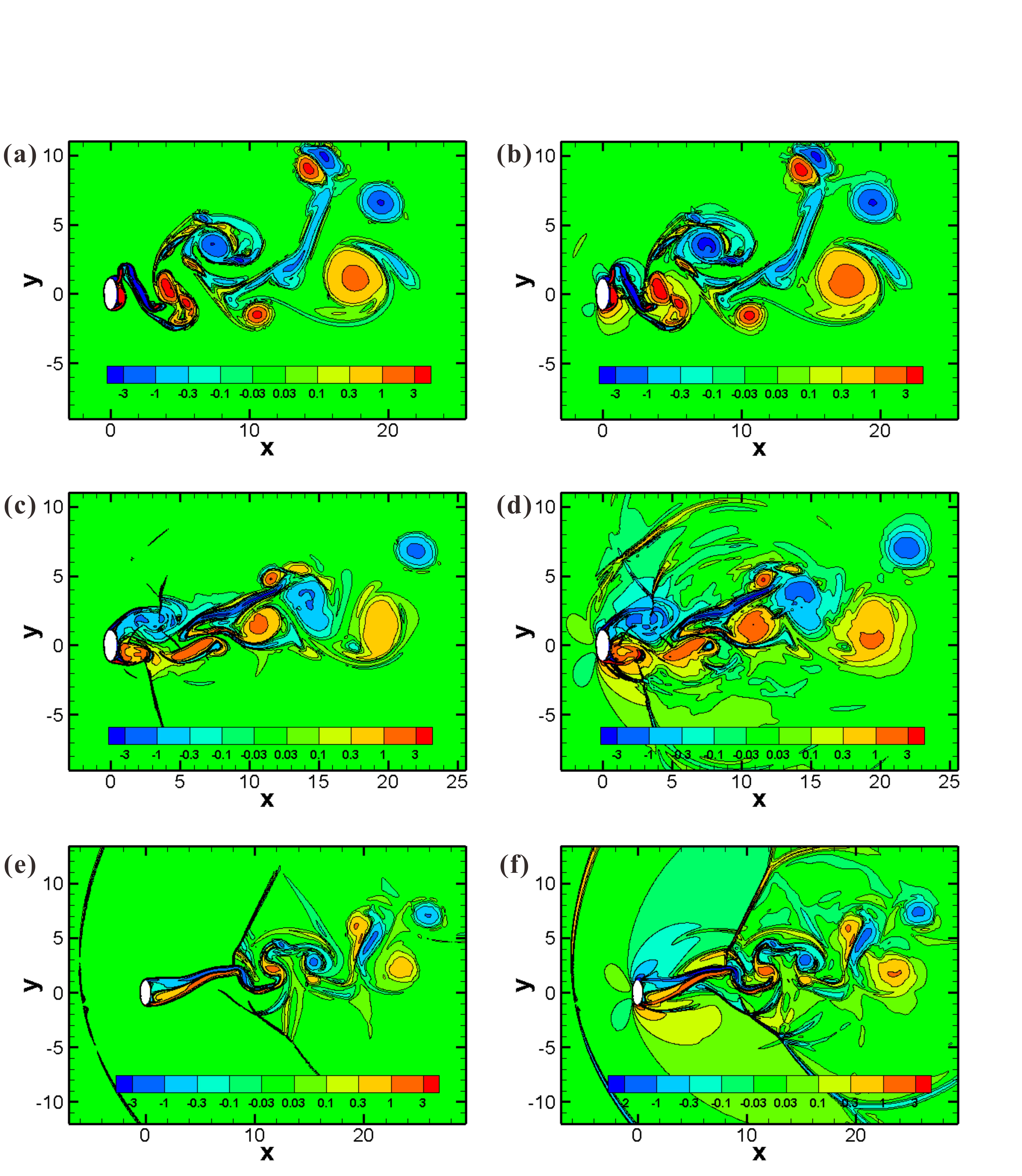}}
  \caption{ Instantaneous $\rho\po$ (a, c, e) and $\po^*$ (b, d, f) contour of compressible flow around an  elliptical airfoil with $Re=1000$ at $t/T=35$ with $M=0.4$ (a, b), $0.8$ (c, d) and $1.2$ (e, f), respectively.}
\label{fig.4-1}
\end{figure}

In Subsection~\ref{sec.3.1}, the finite-domain impulse theory was naturally generalized to compressible flow in terms of $\po^*=\nabla\times(\rho\pu)$. But the dynamic complexity makes it difficult to find good $\Sigma$ satisfying \er{no-cut*} due to the widely spread field of $\nabla\rho\times\pu$ in almost the whole region within the furthest wave. This situation becomes more severe in transonic and supersonic flows as shown in Fig. \ref{fig.4-1}. We found  that at all Mach numbers the distributions of $\rho\po$ and $\po^*$ are quite similar in the wake, but the structures of the latter has larger magnitude and less ``clean'' no-cut region. Besides, the structure of $\om^*$ is clearly seen along the shock. In contrast, the vorticity-moment theory of Subsection~\ref{sec.3.3} in terms of $\rho \om$ performs better.

To check the applicability of minimum-domain compressible vorticity-moment formulation \er{CF6} with assumed existence of good minimum $\Sigma$, we present the results for three Mach numbers: $ M=0.4$, $0.8$ and $1.2$. We chose the $\Sigma$'s that are not too big and cut the least field vorticity. It was found that (figure not shown) whatever the Mach number is, once $\Sigma$ encloses the furthest wave front in the flow field, all formulas behave good including the global-form formula \er{impulse-Huang}. At the opposite extreme, both \er{F*-nocut} and \er{CF6} do not work when $\Sigma$ obviously cuts through the wake structures.

\begin{figure}
  \centerline{\includegraphics[width=0.9\textwidth]{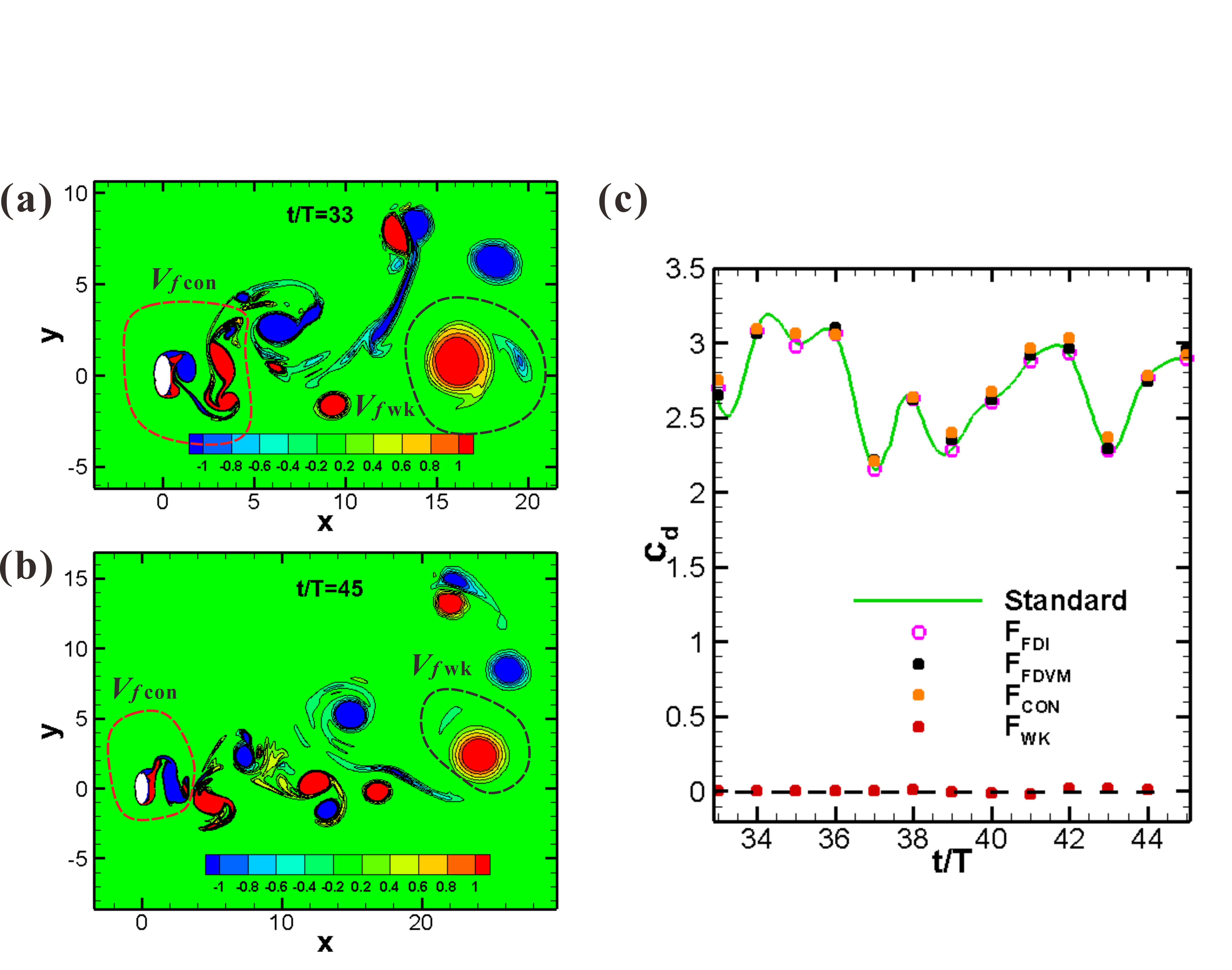}}
  \caption{(a, b) Instantaneous vortical structures of compressible flow with $ M=0.4,\ Re=1000$ around an  elliptical airfoil at $t/T=33$ and $45$, the magnitude of the vortex core in the wake is $O(10)$. (c) The time-dependent drag coefficient $c_d$, calculated by \er{F*} $({\rm F_{FDI}})$, \er{F-rho} $({\rm F_{FDVM}})$, \er{CF6} $({\rm F_{CON}})$, \er{frhoi} $({\rm F_{WK}})$, and \er{standard} ${\rm (Standard)}$.}
\label{fig.4}
\end{figure}

\begin{figure}
  \centerline{\includegraphics[width=0.9\textwidth]{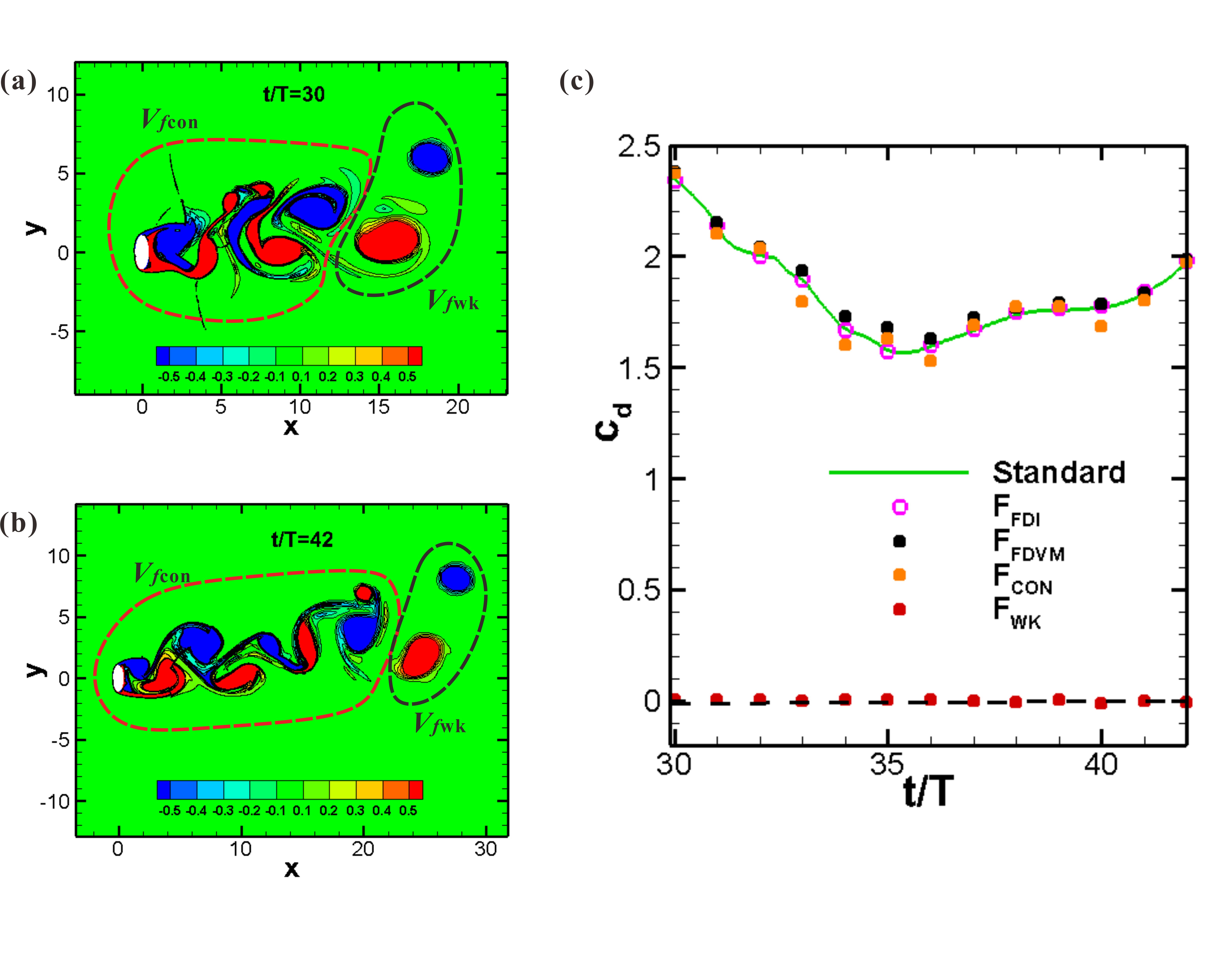}}
  \caption{(a, b) Instantaneous vortical structures of compressible flow with $ M=0.8,\ Re=1000$ around an elliptical airfoil at $t/T=30$ and $42$, respectively. The magnitude of the vortex core in the wake is less than $10$. (c) The time-dependent drag coefficient $c_d$, calculated by \er{F*} $({\rm F_{FDI}})$, \er{F-rho} $({\rm F_{FDVM}})$, \er{CF6} $({\rm F_{CON}})$, \er{frhoi} $({\rm F_{WK}})$, and \er{standard} ${\rm (Standard)}$.}
  \label{fig.5}
\end{figure}

\begin{figure}
  \centerline{\includegraphics[width=0.9\textwidth]{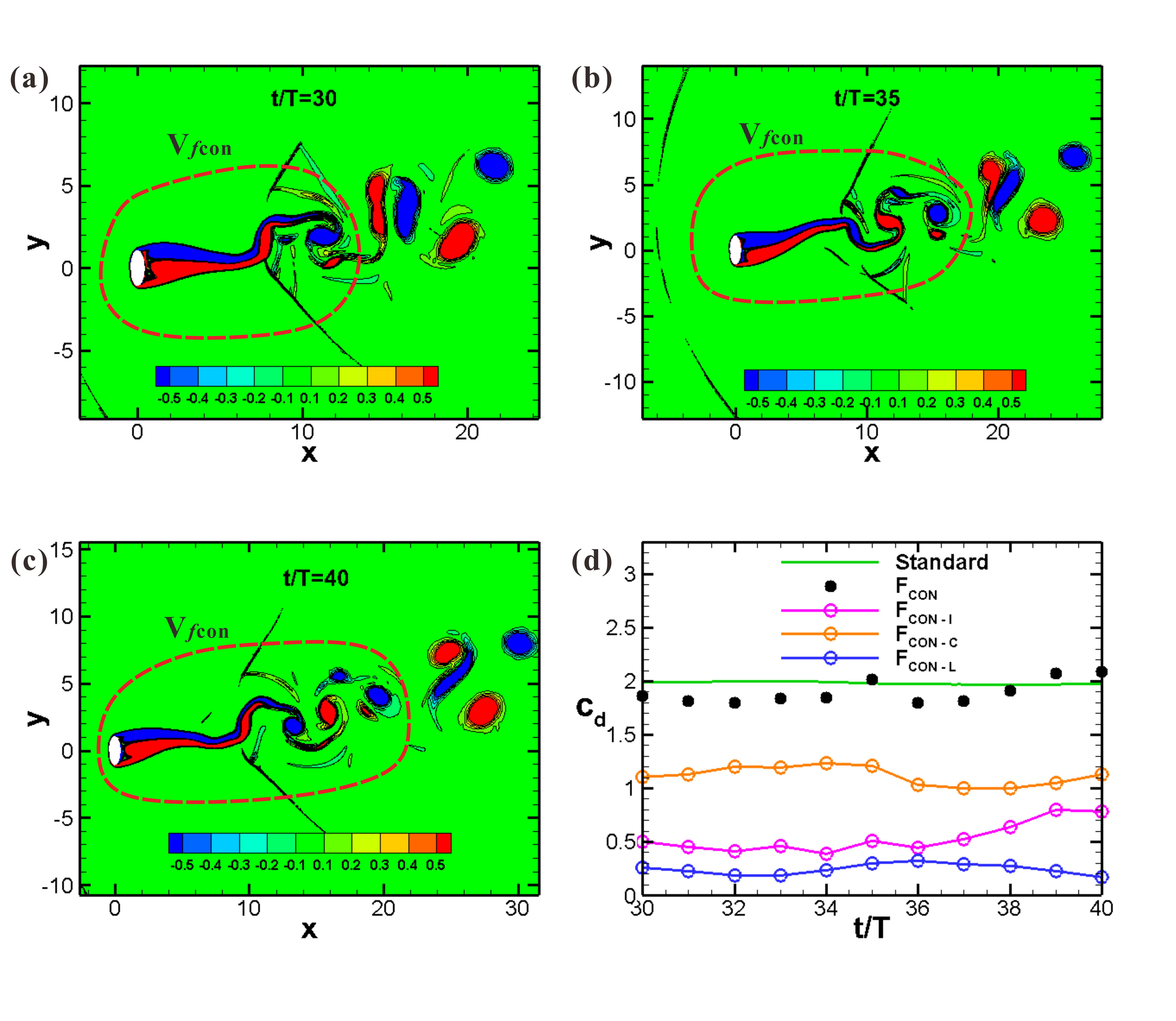}}
  \caption{(a, b, c) Instantaneous vortical structures of compressible flow with $ M=1.2,\ Re=1000$ around an elliptical airfoil at $t/T=30, 35$ and $40$, respectively. The magnitude of the vortex core in the wake is less than $5$. (d) The drag coefficient $c_d$ computed by \er{CF6} $({\rm F_{CON}})$ and \er{standard} $({\rm Standard})$. The line legend marked by $I$, $C$, and $L$ represent the force contributed by $-d\pI_{\rho f{\rm con}}/dt$, the integral of time-rate of $-\px\times \pw_\rho$, and that of $-\rho\po\times \pu$ in \er{CF6}, respectively.}
\label{fig.6}
\end{figure}

Figure \ref{fig.4}(a, b) illustrate the instantaneous contours of vorticity at $t/T=33$ and $45$ with $M=0.4$, where $\cV_{f\rm{wk}}$ follows the same compact vortical structure. Figure \ref{fig.5}(a,b) illustrate the vortical structure at $t/T=30$ and $42$ with $M=0.8$, where $\cV_{f\rm{wk}}$ contains the wake vortices excluded by $\cV_{f\rm{con}}$. Figure \ref{fig.6}(a, b, c) shows the instantaneous vorticity contours at $t/T=30,\ 35$ and $40$.

Figures \ref{fig.4}(c) and \ref{fig.5}(c) show the drag coefficient $c_d$ calculated by different formulas varying with $t/T$. We first observe that in compressible flow \er{frhoi} is valid. Thus \er{CF6} works well at these Mach numbers. Remarkably, we found that (figure not shown) the same reason for the failure of \er{F*-nocut} also makes \er{impulse-Huang} unable to work well even $\Sigma$ is big enough to enclose all strong wake vortices but only excludes some weak field of $\nabla\rho\times\pu$.

When the flow compressibility is enhanced at supersonic flow with $ M=1.2$, the relative error of \er{CF6} is about $10\%$ in force prediction because of inevitably cutting some vorticity, although its respective arbitrary-domain versions still work well. Figure \ref{fig.6} shows that strong shocks not only squeeze the wake vorticity to a narrow region but also suppress the flow unsteadiness.

It seems that as the Mach number increases, the wave-like behavior of compressible flow is getting more dominant, and the wake structures are gradually continuous. Truly good $\Sigma$ satisfying either \er{no-cut} or \er{no-cut*} has to be much larger and is more rarely identifiable. Consequently, even the vorticity-moment theory no longer works well in its minimum-domain form.  But \er{F*-nocut} deteriorates faster than \er{CF6} due to its sensitivity to the wave-like structures caused by $\na \rho\times \pu$. Some detailed comparison of the distributions of $\px\times \po^*$ and $\px\times \rho\po$ (figure not shown) supports this physical interpretation.

It has to be stressed that, however, the Mach number is not the unique parameter to control the prediction ability of those minimum-domain theories. Other flow parameters, such as body geometry and angles of attack, might have important influence. A full consideration of this case-dependent factors would lead to a more objective assessment of compressible minimum-domain theories.

\section{Concluding remarks}\label{sec.5}

The main findings of this paper are the followings:

1. Despite its remarkable neatness and generality, the classic global impulse theory of unsteady aerodynamic force requires knowing the entire field of vorticity or the curl of fluid momentum in the free space, which is beyond the ability of modern CFD and EFD. To relax this requirement, we extend the theory to finite-domain formulation and test the results numerically, for both incompressible and compressible flows. This opens an avenue of combining the theory with advanced numerical-laboratory  experiment in various practical studies. However, while the force formulas can be exactly expressed in an arbitrary finite domain, the mechanism between the force and flow structures is not clear because of the cumbersome $\Sigma$-integrals. Thus, the central task should be simplifying the finite-domain formulation to recover maximally the neatness and generality of classic theory. When the wake structures are discrete and compact, the desired simplification is found indeed feasible.

2. For incompressible flow with discrete wake, it is proven that the total force on the body is solely determined by the body-connected vortical structures. Once shed off, any compact structure in the wake will exert no force to the body apparently. This leads to a minimum-domain impulse theory for discrete wake.

3. For compressible flow, the curl of fluid momentum $\rho \pu$ (which is the only way to express compressible impulse) contains the compressibility effect $\na\rho\times \pu$ that does not have  discrete pattern. Hence a generic minimum-domain impulse theory of practical value does not exist. This fact underscores that the wake discreteness is a feature inherent only in transverse process measured by vorticity. Confining to transverse process alone in compressible flow, it is proved feasible to use the integral of $\px\times \rho \po$ to construct a minimum-domain vorticity-moment theory. But since the density jump across shocks and associated continuous vorticity field behind the shocks, for supersonic flow it is more difficult to find discrete structures of $\rho \po$.

4. In addition to obtaining the minimum-domain theory for flow with discrete wake, which provides convenience to CFD/EFD,
a more important finding of this work is that the key flow structures for producing force are near the body. True, those vortical
structures already disconnected from the body still coexist and
interact with the near field; but their effect on the force can
well be reflected by the minimum-domain Lamb-vector integral. This advantage over the classic global theory is obvious in
the studies of biological locomotion and other unsteady aerodynamics problems.

Finally, we remark that as mentioned in Section~\ref{sec.4}, the use of derivative-moment transformation (DMT) to remove unwanted boundary integrals was merely an artifice instead of a consistent approach. Actually a more thorough application DMT to remove all boundary integrals would lead to the {\it kinetic form} or the so-called diffusion form and boundary form \citep{Wu1993, Wu2015}, which holds for the entire Mach number regime from incompressible to supersonic flows and in which all kinematic effects are filtered out, leaving only causal mechanisms. The analysis domain can then be further reduced and even to body-surface integrals only. For one of the relevant works see \cite{Zou2017}. The neatness of the kinetic form is comparable to the global impulse theory. The application and unique role of such kinetic-form of theory in unsteady aerodynamics problems will be discussed elsewhere.

\section*{Acknowledgements}
\addcontentsline{toc}{section}{Acknowledgements}
This work was partially supported by NSFC (Grant No. 10921202, 11221062, 11521091, 11472016) of China. The authors would like to thank Prof. J. C. Wu for his insightful comments on the first draft of this paper. We are also grateful to Prof. Yi-Peng Shi and Messrs. Shu-Fan Zou and An-Kang Gao for their very valuable discussions. L. Q. gratefully acknowledges the support of the Boya Postdoctoral Fellowship.

\appendix

\section{Derivative-moment-transformation}\label{app.A}
In developing the theory two vectorial integral identities, named derivative-moment-transformation (DMT for short), will be frequently used: in $n$-dimensional space with $n=2,3$, set $k=n-1$, for any piecewise differentiable vector field $\pgg$ and scalar field $\zeta$, there is
\bsubeq\lb{DMT}
\beqn
 \int_V\pgg dV &=& \fr{1}{k}\int_V\px\times (\na\times \pgg)dV -\fr{1}{k}\int_{\pat V}\px\times (\pn\times \pgg)dS,\lb{DMTa}\\
 \int_V\na\zeta dV &=& -\fr{1}{k}\int_{\pat V}\px\times (\pn\times \na\zeta)dS.\lb{DMTb}
\eeqn
\esubeq
Note that these identities are independent of the origin of the position vector $\px$, say $\px_0$, as can be checked by the fact if the operation $\px\times $ is removed then their left-hand sides vanish and righ-thand sides are just the Gauss theorem. Thus, all the new force formulas obtained in this paper derived by DMT must also be $\px_0$-independent, although some individual terms therein may depend on the choice of $\px_0$.

\section{Kinematic content of $d\pS^*_f/dt$}\label{app.B}

For any material control volume $\cV_f$ we have ($\ppm = \rho \pu$)
\beqn\lb{3T}
 k\fr{d \pS^*_f}{dt} &=& \fr{d}{dt} \int_{\pat \cV_f} \px \times (\pn\times \ppm) dS =
 \int_{\pat \cV_f} (\pn \ppm\cdot \pu - \ppm u_n) dS \nonumber  \\
 & & + \int_{\pat \cV_f} \px \times \left (\pn\times \fr{D \ppm}{D t} \right) dS + \int_{\pat \cV_f} \px\times \left(\fr{D}{Dt} d\pS\times \ppm \right).
\eeqn
Of the three terms here, the third term is most complicated. We give detailed algebra,
\beq\lb{T20}
 \fr{1}{dS}\px\times \left(\fr{D}{Dt}d\pS\times \ppm \right) = -[(\pn \times \na) \times \pu] \px \cdot \ppm + \ppm [ (\pn \times \na) \times \pu] \cdot \px.
\eeq
Denote $\pD \equiv \pn \times \na$, then for the first term, there is
\bsubeq
\beqn
 -\ep_{ijk}(D_j v_k)x_l m_l &=& \ep_{ijk}[-D_j(u_k x_l m_l) + u_k m_l (D_j x_l) + u_k x_l (D_j m_l)] \nonumber\\
 &\Rightarrow & \ppm u_n - \pn \ppm \cdot \pu + u_n \na \ppm \cdot \px - \pn \pu \cdot \na \ppm \cdot \px,\lb{T21}
\eeqn
where the arrow indicates neglecting the term of vanishing integral due to the general Stokes theorem, and use was made of a relation
$$D_lx_j = \ep_{lpq}n_p\pat_qx_j =\ep_{lpj}n_p.$$
Then for the second term there is
\beqn
 m_i x_j \ep_{jkl} D_k u_l &=& \ep_{jkl} [D_k (m_i x_j u_l) - m_i u_l (D_k x_j) - u_l x_j (D_k m_i)] \nonumber\\
 &\Rightarrow & (1-n) \ppm u_n - (\px \cdot \na \ppm) u_n + (\pn \cdot \px)\pu\cdot \na \ppm. \lb{T02}
\eeqn
\esubeq
The sum of the two terms is ($\po^* = \na \times \ppm$)
$$ \fr{1}{dS}\px\times \left(\fr{D}{Dt}d\pS\times \ppm \right) \Rightarrow (1-k) \ppm u_n - \pn \ppm \cdot \pu + \px \times \po^* u_n - \px\times [\pn\times (\pu \cdot \na \ppm)].$$
Thus,
\beqn\lb{A2}
 \int_{\pat \cV_f} \px \times \left(\fr{D}{Dt} d\pS \times \ppm \right)
 &=& \int_{\pat \cV_f} \px \times \po^* u_n dS  -\int_{\pat \cV_f} \px\times [\pn\times (\pu \cdot \na \ppm)] dS\nonumber\\
 &-&(k-1) \int_{\pat \cV_f} \ppm u_n dS  - \int_{\pat \cV_f} \pn \ppm \cdot \pu dS.
 \eeqn
Therefore, by substituting \er{A2} into \er{3T}, we obtain
\beq\lb{3T-3-0-1}
 \fr{d \pS^*_f}{dt} = \fr{1}{k}\int_{\pat \cV_f} \px \times \left(\pn\times \ppm_{, t} \right) dS + \fr{1}{k}\int_{\pat \cV_f} \px \times \po^* u_n dS - \int_{\pat \cV_f} \ppm u_n dS.
\eeq
Moreover, recall that $\pat V_f = \pat B + \Sigma$ where $\pat B$ is a material surface and $\Sigma$ is chosen to be a material control surface, and that $\ppm_{,t} = \rho \pa - \na \cdot ( \pu\ppm)$, there is
\beqn\lb{3T-3-0-2}
 \fr{d \pS^*_f}{dt} &= & - \fr{1}{k} \int_{\cV_f} \px \times \left[ \na\times \na \cdot (\pu \ppm) \right] dS + \fr{1}{k}\int_{\pat \cV_f} \px \times \po^* u_n dS \nonumber \\
 & & + \fr{1}{k}\int_{\pat \cV_f} \px \times \left(\pn\times \rho \pa \right) dS.
\eeqn
Note that $\na \cdot ( \pu\ppm) = \rho \pf + \na K$ as mentioned in Section~\ref{sec.3.1}, thus \er{3T-3-0-2} reduces to
\beqn\lb{B7}
 \fr{d \pS^*_f}{dt} &= &-\int_{\cV_f}\pf dV
  - \fr{1}{k} \int_{\pat \cV_f} \px \times \left[ \pn\times (\rho \pf) \right] dS  \nonumber \\ & &+ \fr{1}{k}\int_{\pat \cV_f} \px \times \po^* u_n dS
 + \fr{1}{k}\int_{\pat \cV_f} \px \times \left(\pn\times \rho \pa \right) dS,
\eeqn
which for incompressible flow reads
\beq\lb{B8}
 \fr{d\pS_{f}}{dt} = -\int_{\cV_f}\rho\po\times\pu dV +\fr{1}{k}\int_{\pat \cV_f}\px\times\rho\pu \om_n dS + \fr{1}{k}\int_{\pat \cV_f} \px \times \left(\pn\times  \rho \pa \right) dS.
\eeq

\section{Dynamic content of $d\pI_{\rho f}/dt$}\label{app.C}

Consider the flow in a material volume $\cV_f$, there is
\beq\lb{dIdt-2}
 \fr{d\pI_{\rho f}}{dt}=\fr{1}{k}\int_{\cV_f}\left(\pu\times \rho\po+\px\times\fr{\cD\rho\po}{\cD t}+\px\times \rho\po\vartheta\right)dV,
\eeq
in which,
\beq
\fr{1}{k}\int_{\cV_f}\px\times\fr{\cD\rho\po}{\cD t}dV= \fr{1}{k}\int_{\cV_f}\px\times (\rho\po\cdot\na\pu-2\rho\po\vartheta+\na\ln\rho\times\na p)dV\nonumber+\pJ_\mu,
\eeq
where
\beq\lb{Jmu}
\pJ_\mu\equiv\fr{1}{k}\int_{\pat \cV_f}(\px\times\rho\ps+\pt)dV.
\eeq
Then since for $k=3$ only,
$$\px\times (\rho\po\cdot \na\pu) = \na\cdot (\rho\po\px\times \pu)-\rho\po\times \pu-\px\times\pu(\na\rho\cdot\po),$$
for both $n=2,3$ we obtain
\beq
\fr{d\pI_{\rho f}}{dt}=-\int_{\cV_f}(\rho\po\times\pu +\px\times\pw_\rho) dV+\fr{1}{k}\int_{\pat \cV_f}\px\times\pu \rho\omega_n dS+\pJ_\mu.
\eeq
with $\pw_\rho$ being defined as
$$k\pw_\rho\equiv\rho\po\vartheta+\pu\na\rho\cdot\po+\rho\na T\times\na S$$
In incompressible flow, it is
\beq\lb{A4}
 \fr{d\pI_f}{dt} = - \int_{\cV_f}\rho\po\times\pu dV +\fr{1}{k}\int_{\pat \cV_f}\px\times\pu \rho\omega_n dS+\pJ_\mu.
\eeq

%

\end{document}